\documentclass[conference]{IEEEtran}
\IEEEoverridecommandlockouts
\usepackage{cite}
\usepackage{amsmath,amssymb,amsfonts}
\usepackage{algorithmic}
\usepackage{graphicx}
\usepackage{textcomp}
\usepackage{xcolor}
\def\BibTeX{{\rm B\kern-.05em{\sc i\kern-.025em b}\kern-.08em
    T\kern-.1667em\lower.7ex\hbox{E}\kern-.125emX}}
\begin{document}

\title{SBAN: A Framework \& Multi-Dimensional Dataset for Large Language Model Pre-Training and Software Code Mining
 \\
{\footnotesize \textsuperscript{*}SBAN: LLMs and Code Mining}
\thanks{ }
}

\author{
\IEEEauthorblockN{Hamed Jelodar, Mohammad Meymani, Samita Bai, and Roozbeh Razavi-Far, Ali A. Ghorbani}
\IEEEauthorblockA{\textit{Canadian Institute for Cybersecurity, Faculty of Computer Science} \\
\textit{University of New Brunswick}, Fredericton, Canada \\
Email: \{h.jelodar, mohammad.meymani79, samita.bai, roozbeh.razavi-far, ghorbani\}@unb.ca}
}

\maketitle

\begin{abstract}
  This paper introduces SBAN (Source code, Binary, Assembly, and Natural Language Description), a large-scale, multi-dimensional dataset designed to advance the pre-training and evaluation of large language models (LLMs) for software code analysis. SBAN comprises more than 3 million samples, including 2.9 million benign and 672,000 malware respectively, each represented across four complementary layers: binary code, assembly instructions, natural language descriptions, and source code. This unique multimodal structure enables research on cross-representation learning, semantic understanding of software, and automated malware detection. Beyond security applications, SBAN supports broader tasks such as code translation, code explanation, and other software mining tasks involving heterogeneous data. It is particularly suited for scalable training of deep models, including transformers and other LLM architectures. By bridging low-level machine representations and high-level human semantics, SBAN provides a robust foundation for building intelligent systems that reason about code. We believe that this dataset opens new opportunities for mining software behavior, improving security analytics, and enhancing LLM capabilities in pre-training and fine-tuning tasks for software code mining. 
%In this paper, we introduce SBAN (Source, Binary, Assembly, and Natural Language Description), a large-scale, multi-dimensional dataset designed to advance the pre-training and evaluation of large language models (LLMs) for software code analysis. SBAN comprises over 3.3 million samples, including 2.8 million benign and 541,000 malware instances, each represented across four complementary layers: binary code, assembly instructions, natural language descriptions, and source code. This unique multi-modal structure enables research in cross-representation learning, semantic understanding of software, and automated malware detection. SBAN is particularly suited for data mining tasks involving heterogeneous data and supports scalable training of deep models, including transformers and other LLM architectures. By bridging low-level machine representations and high-level human semantics, SBAN provides a robust foundation for building intelligent systems that reason about code. This dataset opens new opportunities for mining software behavior, improving security analytics, and enhancing LLM capabilities in program understanding.\\
\end{abstract}


\section{Introduction}

The rapid growth of software development and the increasing complexity of modern applications have heightened the importance of effective software code mining \cite{b1}. Mining software, including source code, binaries, and their natural language descriptions, is crucial for numerous tasks such as malware detection, program analysis, and automated code summarization \cite{b2}. However, constructing large-scale, high-quality datasets that capture the multifaceted nature of software remains a significant challenge. Existing datasets often focus on a single representation layer, limiting the scope of cross-modal learning and comprehensive analysis \cite{b3}. This fragmentation hinders the development of advanced machine learning models, particularly LLMs, which benefit from diverse and aligned data sources for pre-training and fine-tuning.

\begin{figure}[h!]
\centering
\includegraphics[width=85mm]{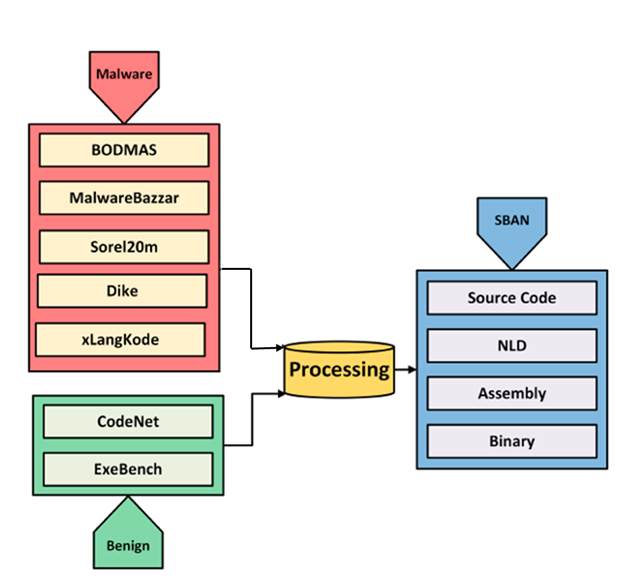}
\caption{ Relation between different sources in the SBAN dataset.}
\label{fig:method}
\end{figure}

Previous efforts have provided valuable resources for specific software representations, such as source code repositories or binary malware datasets \cite{b5}, but rarely integrate multiple software layers within a single benchmark \cite{b4}. This gap restricts the potential to leverage rich semantic relationships across different code modalities and hinders progress in unified code understanding and analysis.\\

To address these challenges, we present SBAN (Source, Binary, Assembly, and Natural Language Description), a multi-dimensional dataset comprising over 3.3 million real-world software samples that encompass four distinct but interconnected representation layers: 1) source code, 2) binary, 3) assembly representations, and 4) natural language descriptions, as showed in the Figure 1. To the best of our knowledge, SBAN is the first dataset to integrate these layers into a unified benchmark, enabling rich cross-modal code mining and analysis. It is specifically designed to support the pre-training and evaluation of large language models for diverse software tasks, including code understanding, malware detection, and program analysis.

\subsection*{ Research Contributions}

Comprehensive software analysis requires datasets that integrate multiple code representations to enable effective cross-modal learning. SBAN addresses this need by providing a large-scale, four-dimensional dataset to advance LLM-based software mining and security research. This work makes the following key contributions:

\begin{itemize}
    \item We introduce \textbf{SBAN}, a large-scale, multi-dimensional dataset with over 3 million samples, covering binary, assembly, source code, and natural language descriptions of both benign and malware software.

    \item To the best of our knowledge, SBAN is the first dataset to integrate these four distinct software representation layers in a unified benchmark, enabling rich cross-modal code analysis.

    \item SBAN is specifically designed to support pre-training and evaluation of large language models (LLMs) for tasks such as code understanding, malware detection, and program analysis.

    \item The dataset supports diverse data mining applications, including malware classification, code summarization, anomaly detection, and deep semantic code mining. It is available at https://www.unb.ca/cic/datasets/sban-dataset-2025.html .

\end{itemize}

\section{Related work}

Due to the increasing sophistication of malware attacks and the growing risk of malware generation through LLMs, there is a clear need for research focused on the development of malware datasets for pre-training and fine-tuning LLMs. Various researchers have proposed approaches for utilizing diverse malware datasets to train and adapt LLMs for cybersecurity tasks.

Recently, Jiang et al. \cite{b10} introduced Nova, a generative LLM tailored for assembly code analysis, addressing challenges such as low information density and diverse compiler optimizations. The authors employ extensive open-source C/C++ codebases, including “The Stack” and “AnghaBench.” By compiling the code at various optimization levels (O0 to O3), they generated approximately 4.3 million assembly functions. Nova employs a hierarchical attention architecture that captures semantics across tokens, instructions, and entire functions. To ensure robustness, it leverages a contrastive learning objective that encourages representations to remain consistent across different compiler optimizations. Following pre-training on this large corpus, Nova was fine-tuned for downstream tasks such as binary similarity detection and binary decompilation, aiming to recover high-level source code from assembly.

In a related study, Li et al. \cite{b11} aimed to learn general-purpose embeddings for binary assembly instructions through pre-training for language models. They created a large, unlabeled corpus by compiling 3,266 benign software programs, including multiple versions of GNU Binutils, Coreutils, Diffutils, and Findutils, using various compiler optimization levels. This process generated approximately 2.25 billion assembly instructions. On this data, they pretrained a BERT-style model called PalmTree using self-supervised objectives such as masked token prediction and next-instruction prediction. PalmTree effectively captures the structural nuances of assembly language, including opcodes, operands, and addressing modes, outperforming prior embedding approaches. The study demonstrates that pre-training on binary code corpora can enhance downstream tools for malware detection and vulnerability analysis.

Harang and Rudd \cite{b12} introduced SOREL-20M, a large-scale benchmark dataset designed for machine learning applications in malware analysis. It contains 20 million Windows PE files equally split between malware and benign samples, each annotated with extensive metadata and pre-extracted static features such as header information, imported functions, strings, and byte histograms. The dataset includes high-confidence labels derived from consensus among multiple antivirus engines. Notably, it also offers approximately 10 million “disarmed” malware binary files that retain executable structure but have been stripped of harmful functionality, making them safe for research and model training. Although SOREL-20M is not a traditional code pre-training corpus, it is widely used to develop and benchmark deep learning models for malware detection at scale. The authors also provide baseline models (e.g., neural networks and gradient-boosted trees) and open-source tooling to facilitate community experimentation and advancement.

Demirci and Acarturk \cite{b13} framed the disassembled malware assembly code as a language corpus suitable for static analysis. They extracted instructions from the \textit{.text} sections of approximately 822 malicious and 672 benign Windows PE files. The GPT-2 model was pretrained on the unlabeled assembly instructions to capture the underlying "language" of malware. This model was then integrated with a stacked BiLSTM network for classification. Their method achieved strong performance in distinguishing between benign and malicious code by effectively modeling the syntactic and semantic relationships among instructions. The study highlights the potential of applying NLP techniques such as transformers and language modeling to malware detection, particularly in learning informative instruction embeddings that can handle challenges like code obfuscation or packing.

\begin{figure*}[htbp]
  \centering
\includegraphics[width=99mm]{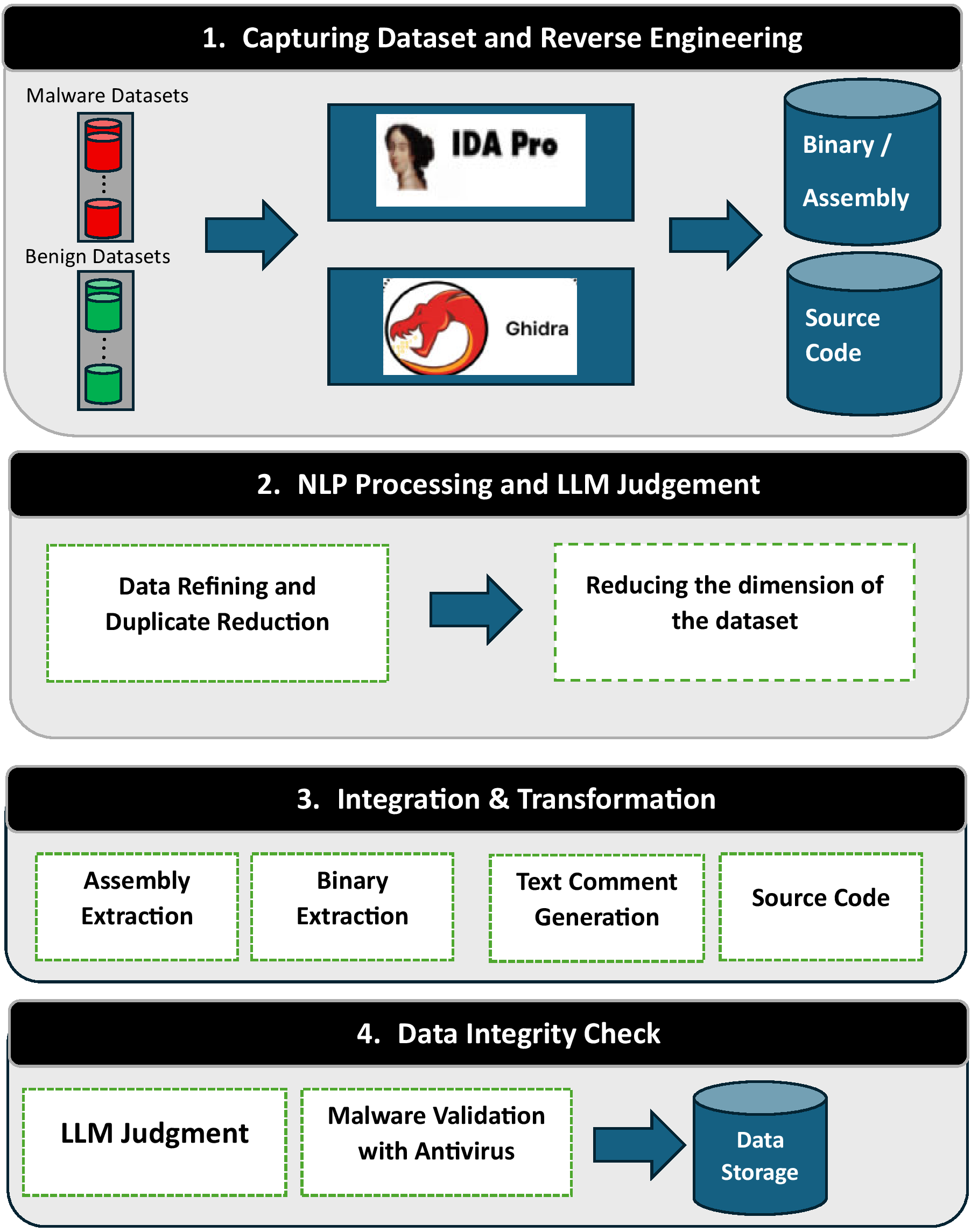}
\caption{A view of SBAN dataset pipline.}
\label{fig}
\end{figure*}

Anderson and Roth \cite{b14} introduced the EMBER dataset in 2018 as one of the first large-scale, publicly available resources for static malware detection. It contains feature vectors for approximately 1.1 million Windows PE files, equally split between malicious and benign samples—collected during 2017–2018. Instead of raw binaries, EMBER provides structured JSON-based features such as header metadata, imported APIs, section entropy, string statistics, and byte histograms, enabling easier model training. The authors released a baseline LightGBM model, showing that traditional feature-based methods were highly competitive at the time.
Although EMBER was not designed for language modeling, it has since been leveraged in LLM research. Structured feature sequences from EMBER have been used to fine-tune transformer-based models like CodeBERT for malware classification tasks. 

The studies discussed above collectively demonstrate the growing interest in leveraging large-scale malware datasets for pre-training and fine-tuning language models. While most prior work focuses on traditional ML or representation learning, recent efforts highlight the potential of LLMs to capture complex patterns within binary and assembly code. Building on these insights, our work aims to further explore and enhance LLM-based approaches for understanding the understanding and detection of benign and malware samples.Figure 2 show a general view of data-collection pipline to prepare the SBAN dataset.

\section{SBAN Dataset}
The proposed dataset, \textbf{SBAN}, is a large-scale, multi-dimensional collection of software samples designed for comprehensive code analysis and large language model pre-training. It includes over 3.3 million samples, comprising 2.8 million benign and 541,000 malware instances.

\section{Dataset Architecture}

The SBAN dataset is structured around four interconnected representation layers, each providing unique and complementary perspectives on code samples:

\begin{itemize}
    \item \textbf{Binary Code:} Raw executable files capturing low-level machine instructions, essential for understanding software behavior and malware analysis at the hardware interaction level.

    \item \textbf{Assembly Code:} Disassembled instructions that provide a human-readable, low-level abstraction of binary executables, facilitating detailed static analysis and reverse engineering.

    \item \textbf{Source Code:} High-level programming language code that reveals the original logic, structure, and intent behind the software, supporting semantic understanding and code summarization tasks.

    \item \textbf{Natural Language Descriptions:} Human-readable explanations and summaries of the software functionality or code segments, enabling bridging between code and natural language processing for advanced language model training.
\end{itemize}

Each sample in SBAN maintains alignment across these four layers, allowing for cross-modal learning and analysis, as showed in Figure 3. This multi-dimensional architecture supports diverse research objectives, including malware detection, code comprehension, and pre-training of LLMs.

\section{Dataset Pipeline}
Figure 2 shows all the phases considered in preparing the dataset. The pipeline begins with the collection of raw code samples, including both benign and malicious binaries from various trusted sources. These binaries are then subjected to a decompilation process, which extracts corresponding source-like code representations suitable for further analysis.\\

Table \ref{tab:M-datasets-size} shows the number of source codes, natural language descriptions, assembly, and binary codes, separately for each malware dataset.  Also, Table \ref{tab:B-datasets-size} shows the number of source codes, natural language description, assembly, and binary codes, separately for each benign dataset.

\begin{table}[h]
    \centering
     \caption{Number of Malware Samples }
    \resizebox{\linewidth}{!}{
    \begin{tabular}{l c c c c}
    \hline
       Dataset  &  Source & NLD & Assem & Binary\\
       \hline
       1. BODMAS & 93711 & 93711 & 92317 & 88605\\
       2. MalwareBazzar & 14746 & 14746 & 14051 & 13973\\
       3. Sorel20m & 81584 & 81584 & 81177 & 79166\\
       4. Dike & 17431 & 17431 & 12138 & 11726\\
       5. xLangKode & 468679 & 468679 & 5974 & 13299\\
       \hline
       Total & 676151 & 676151 & 205657 & 206769\\
       \hline
    \end{tabular}}
    \label{tab:M-datasets-size}
\end{table}

\begin{table}[h]
    \centering
     \caption{Number of Benign Samples }
    \resizebox{\linewidth}{!}{
    \begin{tabular}{l c c c c}
    \hline
       Dataset  &  Source & NLD & Assem & Binary\\
       \hline
       1. CodeNet &152853&152853&152818&152819\\
       2. ExeBench &2800000&2800000&2800000&2800000\\
       \hline
       Total &2952853&2952853&2952818&2952819\\
       \hline
    \end{tabular}}
    \label{tab:B-datasets-size}
\end{table}

\subsection{Captions Data from Different Sources}

To build a robust and diverse dataset for training and evaluating our code captioning model, we collected samples from both benign and malicious code sources, ensuring broad coverage in terms of syntax, semantics, and intent.

\begin{itemize}
    \item \textbf{Benign Samples:} Our benign corpus was meticulously curated from premier datasets such as ExeBench\cite{b6} and CodeNet\cite{b7}, with a targeted focus on C and C++ to capture representative, high-fidelity clean code exemplars. These samples reflect a wide range of real-world software development scenarios, including algorithmic benchmarks, system-level programming, and embedded application code, thereby providing a robust foundation for modeling benign behavior.

 \item \textbf{Malware Samples:} The malware collection integrates diverse and authoritative repositories, including MalwareBazaar \cite{b15}, BODMAS\cite{b8}, DIKE \cite{b16}, xLangKode\footnote{The xLangKode dataset iss developed by the authors of this paper; however, due to space limitations, we are unable to provide further details within this publication.}, and SOREL-20M\cite{b9}, ensuring comprehensive coverage across a wide range of malware families and sophisticated threat vectors. This dataset includes both source-level and binary-level artifacts, encompassing obfuscated code, polymorphic patterns, and advanced persistence mechanisms. Special attention was given to temporal diversity and platform variance, allowing the dataset to represent both legacy threats and emerging malware strains. 
\end{itemize}

\begin{figure*}[htbp]

\includegraphics[width=16cm]{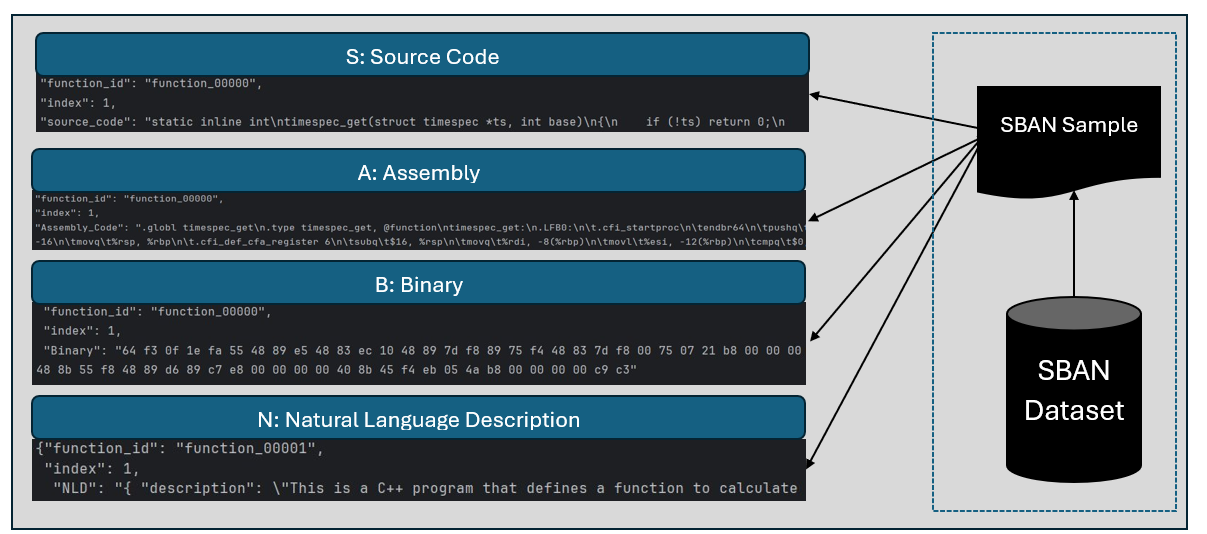}
\caption{ A view of a single sample from the SBAN.}
\centering
\label{fig}
\end{figure*}

The combined dataset was further annotated with behavioral and structural metadata, including function-level summaries, API call traces, and control flow characteristics. This enriched corpus underpins our captioning task, enabling the generation of semantically meaningful and security-aware descriptions of code functionality, both benign and malicious.\\

 \subsubsection{Reverse Engineering and de-compiling }

Some datasets, such as MalwareBazaar\cite{b23}, BODMAS \cite{b21}, and DIKE \cite{b22}, do not include source code as part of their publicly available resources. To address this limitation, we leveraged reverse engineering frameworks such as IDA Pro and Ghidra to decompile the binary files (e.g., .exe or .dll) into approximate C-like code. This allowed us to extract meaningful features and representations from the decompiled samples, which were then used for further analysis and annotation within our dataset pipeline.

 \subsection{NLP pre-processing  and LLM Judment}

\subsubsection{Framework for Evaluating Decompiled Files}

After decompiling the files, we adopted a hybrid framework designed to determine whether each sample is malware or benign. This framework integrates both static code analysis and semantic similarity evaluation, as shown in the Figure 4. Specifically, we compare the decompiled code to the original source code using natural language processing (NLP) techniques.\\

To assess textual similarity, we employed the Sentence-BERT (SBERT) model \cite{b17}, a powerful transformer-based model optimized for capturing semantic similarity between sentences or code comments. SBERT allows us to compute embeddings for both the original and decompiled code representations and then measure their cosine similarity. This approach not only aids in verifying the quality of the decompiled files but also serves as an auxiliary method to flag discrepancies between benign and malicious code samples.\\

\subsubsection{NLP Preprocessing and Code Mining}
One of the most important initial steps in this kind of task is NLP preprocessing. It plays a critical role in refining the data, cleaning noisy samples, and reducing the overall size of the dataset while preserving key information. NLP preprocessing includes tokenization, normalization, removal of duplicates, and abstraction of identifiers.\\

In our workflow, we also employed a LLM, specifically Qwen \cite{b18}, for code refactoring and reuse detection. Qwen has shown high performance in identifying semantically equivalent code blocks and transforming them into standardized formats. Compared to other models, Qwen provides better generalization and code understanding, which is crucial when working with diverse samples from both benign and malware datasets.\\
 
\subsection{Integration and Transformation}

In this phase, we began by preparing the dataset using the Raw Hexadecimal Representation (RHR) of the executable files. This step ensured a consistent low-level format of the binary data, preserving instruction-level details necessary for accurate reverse engineering. The RHR allowed us to standardize input formats for subsequent analysis stages.

Following this, we performed assembly extraction to translate the raw hexadecimal data into human-readable assembly code, enabling us to examine low-level program logic. Finally, we took advantage of the \textbf{Qwen2.5-Coder-1.5B-Instruct} \cite{b18} model for comment generation.

\subsection{Data Integrity Check}

For benign samples, we did not face significant challenges in validation, as the source code was readily available and could be directly verified. However, for malware samples, the validation process was more complex since we had to decompile the functions from executable files. To ensure the integrity and reliability of these samples, we employed a hybrid validation strategy that included: 

1) \textbf{LLM Judgement} – We utilized large language models to semantically assess the quality and coherence of the decompiled code, ensuring that it reflected legitimate functionality and structure, as shown in Figure 4. In this framework, decompiled functions are first evaluated with an LLM model and then compared with the original source code using the SBERT model to understand how similar the decompiled functions are to the original ones.

\begin{figure}[h!]
\centering
\includegraphics[width=85mm]{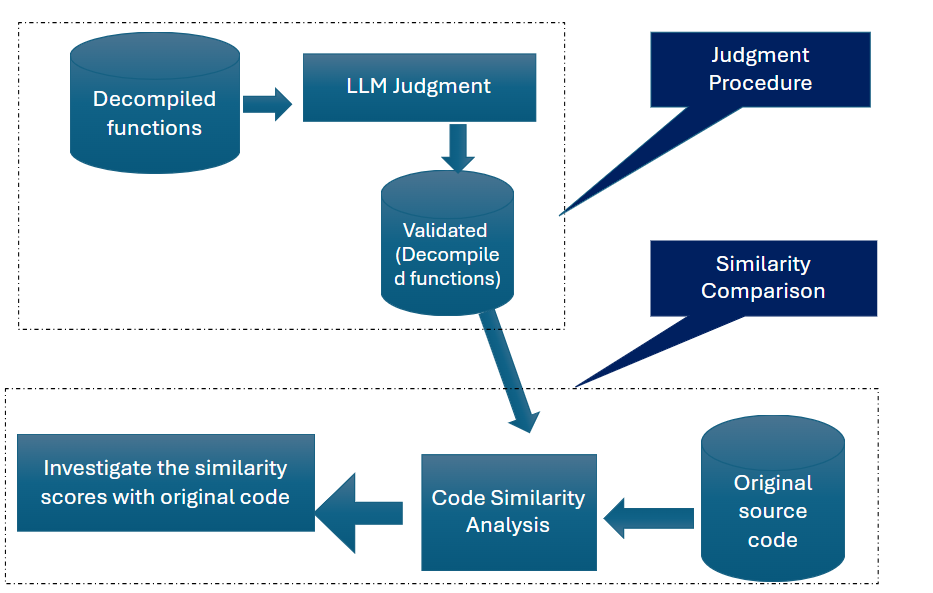}
\caption{Framework for evaluating Decompiled Files}
\label{fig:method}
\end{figure}

2) \textbf{Antivirus Checking} – We cross-validated the malware samples using multiple antivirus engines such as 1) Bitdefender, 2) Kaspersky to confirm their classification and detect any anomalies or false positives.
This dual-layered approach helped us establish a high-confidence dataset for both benign and malicious code samples, forming a solid foundation for downstream analysis.\\

\begin{figure*}[htbp]

\includegraphics[width=18cm]{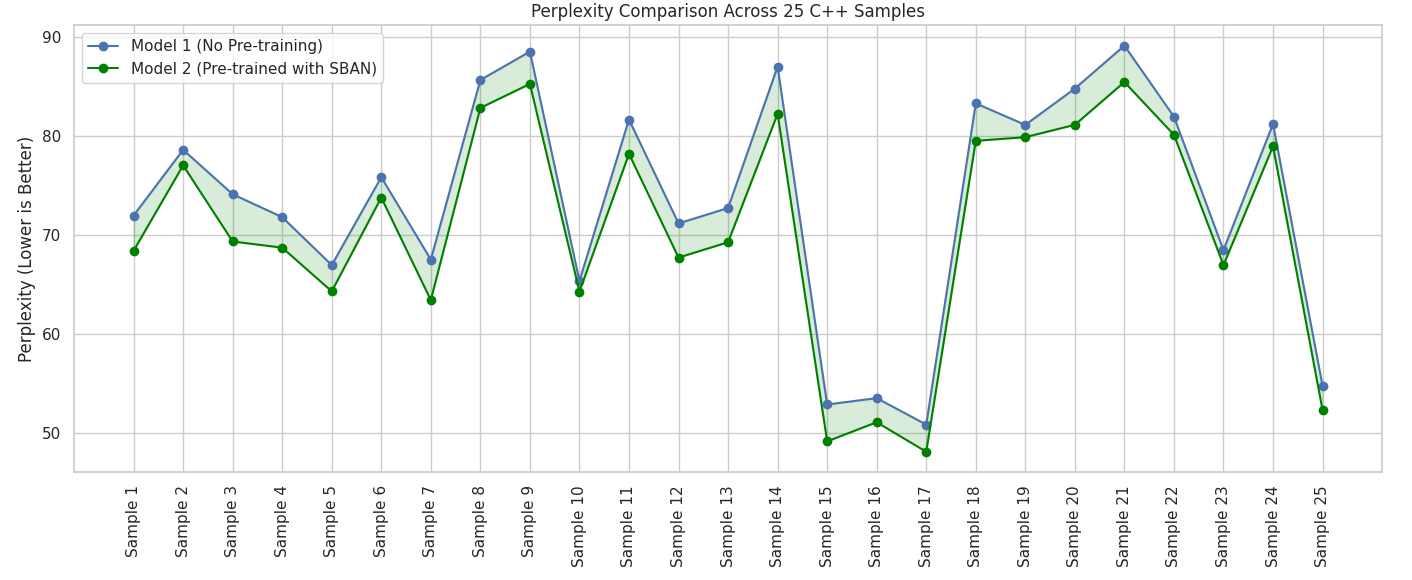}
\caption{Distribution of Perplexity Scores for both Models on Benign Samples}
\centering
\label{fig}
\end{figure*}

\section{Experiments}
\subsection{Computing Infrastructure}
We utilized the Nvidia H100 GPU for pre-training the model. To optimize both computational efficiency and model performance, we incorporated mixed-precision training to improve the performance of the LLM model.

\subsection{Training Settings}
Since the pre-training task for the LLM is unsupervised, there is no need for a training/testing split. The model learns from the entire corpus without task-specific labels on 2.8 million C++ code samples. Figure 4 shows a sample of the SBAN dataset.

\subsection{LLM hyper-parameters}
In this experiment, the hyperparameter settings are: learning rate (2e-5), num\_train\_epochs (3), batch size (128 per device for both training and evaluation), warmup\_steps (500), weight\_decay (0.01), gradient\_accumulation\_steps (1), and evaluation\_strategy ("epoch").

\subsection{LLM pre-training Evaluation}
To demonstrate the practical application of our dataset, we focused on the benign portion for a code mask prediction task. Specifically, we utilized the Qwen/Qwen2.5-Coder-1.5B-Instruct model \cite{b18} and evaluated its performance before and after pre-training on our domain-specific data. The goal was to assess whether targeted pre-training on structured and clean code samples improves the model's ability to understand and predict masked tokens in C++ code.\\

Figure 5 illustrates the distribution of perplexity scores for both model versions, highlighting the impact of pre-training. Model 1, representing the Qwen model without any pre-training, serves as the baseline. It exhibits a wider and more variable range of perplexity values across diverse C++ code samples. This variability suggests inconsistent token prediction confidence, particularly in the presence of long sequences or padded inputs. The model’s unfamiliarity with domain-specific programming patterns is reflected in these fluctuations.

In contrast, Model 2, which underwent pre-training on benign samples from the SBAN dataset, shows a significantly narrower and lower perplexity distribution. This improvement indicates increased prediction stability and better alignment with the syntax and semantics of C++ code. By pre-training on clean, non-malicious samples, the model was exposed to standardized programming structures, reducing semantic noise and enhancing its ability to generalize across coding scenarios.

\section{Limitations and Broader Impact}
While SBAN offers a rich, multi-dimensional dataset for code and malware analysis, certain limitations remain. First, although the dataset spans a wide range of benign and malicious samples, it may still underrepresent rare or emerging malware families. Additionally, the natural language descriptions are generated using large language models, which may occasionally introduce inaccuracies or hallucinations in code explanations.\\

The broader impact of SBAN is promising, yet it calls for responsible use. By enabling large-scale pre-training of LLMs on executable code and software artifacts, SBAN supports research in program understanding, malware detection, and secure software engineering. We also believe there is still room for improvement to increase the diversity of samples in different programming languages, along with their associated binaries or assembly code, as part of future work.

\section{Potential Applications: Where can this dataset be used? }

\subsection{Cybersecurity and Code Intelligence}
The SBAN dataset provides a rich, multi-modal resource for advancing research in cybersecurity, software engineering, and data mining. Its unique alignment of source code, binary, assembly, and natural language description enables a variety of applications, such as malware detection, static code analysis, reverse engineering, and code summarization. Researchers can take advantage of SBAN to train models that detect and explain malicious behaviors, classify software based on its low-level structure, or map between natural language and program logic\cite{b24}. 

\subsection{Multimodal Learning and Software Mining}
Beyond security-focused tasks, SBAN supports broader applications in multimodal learning and software mining. It enables cross-representation learning, code search, and anomaly detection across different levels of abstraction. For example, it can be used to generate natural language summaries of binaries or to identify code similarities using both syntactic and semantic cues. 

\subsection{Benchmarking, pre-training, and Fine-Tuning}
The SBAN dataset serves as a valuable benchmark for evaluating a wide range of machine learning and data mining algorithms across multiple modalities. Its structured, aligned representations allow researchers to assess the performance of models in tasks such as code summarization \cite{b25}, malware classification, and cross-modal translation. By offering consistent mappings between source code, binaries, assembly instructions, and natural language, SBAN facilitates the development and testing of models that require a comprehensive understanding of software artifacts.

\section{Conclusion}
We introduced a novel multimodal dataset to advance data mining applications in LLMs and software code mining. This dataset can be useful for LLM tasks such as pre-training and fine-tuning. We also believe that SBAN has practical applications for both researchers and industry professionals, serving as a valuable resource for improving model performance, security analysis, and automated code understanding across diverse domains.

\end{document}